\documentclass[onecolumn,showpacs,showkeys,preprintnumbers,amsmath,amssymb,superscriptaddress,floatfix,aps]{revtex4}

\usepackage{graphicx}
\usepackage{dcolumn}
\usepackage{bm}

\begin{document}


\title{One-dimensional counterion gas between charged surfaces: \\
Exact results compared with weak- and strong-coupling analysis}

\author{David S. Dean}

\affiliation{ Universit\'e de Toulouse; UPS; Laboratoire de Physique Th\'eorique (IRSAMC);  F-31062 Toulouse, France}
\affiliation{Kavli Institute of Theoretical Physics, University of California, Santa Barbara, CA 93106, USA}

\author{Ron R. Horgan}

\affiliation{DAMTP, CMS, University of Cambridge, Cambridge,  CB3 0WA, UK}
\affiliation{Kavli Institute of Theoretical Physics, University of California, Santa Barbara, CA 93106, USA}

\author{Ali Naji}

\affiliation{Dept. of Physics, Dept. of Chemistry and Biochemistry, and Materials Research Laboratory, University of California, Santa Barbara, 
CA 93106, USA}
\affiliation{Kavli Institute of Theoretical Physics, University of California, Santa Barbara, CA 93106, USA}

\author{Rudolf Podgornik}

\affiliation{Dept. of Physics, Faculty of Mathematics and Physics, University of Ljubljana, and Dept. of Theoretical Physics, J. Stefan
Institute, SI-1000 Ljubljana, Slovenia, and  Lab. of Physical and Structural Biology, National Institutes of Health, MD 20892, USA}
\affiliation{Kavli Institute of Theoretical Physics, University of California, Santa Barbara, CA 93106, USA}

\date{8 January 2009}

\begin{abstract}

We evaluate {\sl exactly} the statistical integral for an inhomogeneous one-dimensional counterion-only Coulomb gas 
between two charged boundaries and from this compute the effective interaction, or disjoining pressure, between the 
bounding surfaces. Our exact results are compared with the limiting cases of weak and strong coupling which are the 
same for 1D and 3D systems.  For systems with a large number of counterions it is found that the weak coupling 
(mean-field)  approximation for the disjoining pressure works perfectly and that fluctuations around the mean-field in 
1D are much smaller than in 3D. In the case of few counterions it works less well and strong coupling approximation 
performs much better as it takes into account properly the discreteness of the counterion charges.

\end{abstract}


\maketitle
\section{Introduction}

Electrostatic effects are predominant in a variety of soft condensed matter systems such as polymers in solutions, 
colloidal suspensions and much of the physics of membranes and films \cite{intro1,intro2}.  The statistical mechanics 
of these problems cannot in general be solved explicitly; however a number of approaches have been devised, starting 
with the mean-field Poisson-Boltzmann equation, that allow for an approximate solution to the problem \cite{hoda}. One 
can improve the accuracy of mean-field calculations by looking at the fluctuations about the Poisson-Boltzmann 
solution, and in a number of cases these fluctuation effects are essential to capture the quantitative behavior of the 
systems under study. The approach of this type can be shown to be valid in the so called weak-coupling regime. More 
recently the strong-coupling expansion, in its essence similar to the virial expansion, has been developed and 
successfully applied to a number of interesting problems \cite{hoda,Naji}. Despite the success of these approaches 
many questions remain to be answered about the intermediate regime and how the cross-over between the weak- and 
strong-coupling limits occurs. Here we show that the one-component 1D inhomogeneous Coulomb fluid can be analytically 
solved and thus provides an ideal test-bed to study the domains of validity of these commonly used approximation 
schemes. 

For a system of two charged surfaces interacting in an electrolyte solution, the effective coupling strength may 
depend on the inter-plate separation, and so the calculation of the force between the plates requires an understanding 
of the both weak and strong coupling regimes and the transition between them.  In two important papers by Lenard 
\cite{len61} and Edwards and Lenard \cite{ed62} a complete solution of the thermodynamics of the two component Coulomb 
gas in one dimension was given. The second of these papers was based on the fact that the underlying Sine-Gordon 
theory for this system can be mapped onto quantum mechanics in one dimension and the thermodynamics can be solved in 
terms of the ground state energy of the corresponding Schr\"odinger equation. Subsequently the system was studied in 
the presence of an electric field and it was shown that it always behaves as a dielectric because of a dimerization of 
the individual charge carriers \cite{ape79}. When the system is not in an electroneutral state, {\em i.e.} when the 
surface charges corresponding to the applied field are not exactly compensated by the counterions of the system, the 
physics of the system is subtly different and confinement phenomena between oppositely charged particles appear 
\cite{aiz80,aiz81}. This confinement is directly related to the fact that the system is dielectric and non-conducting. 
More recently, the one-dimensional Coulomb gas thermodynamics has been studied taking into account additional 
hard-core and dipolar interactions \cite{ver90}. The model has also been studied in the finite size setting of a soap 
film type model where charge regulation mechanism model of the surface, in the presence of salt solution (ions and 
counterions), is included \cite{de98}.

In the present paper we study a related model to those outlined above:  a one-dimensional model of counterions 
confined between two plates, carrying fixed but not necessarily equal charges. This situation is quite different to 
most of those studied above as most of them have concentrated on symmetric electrolyte systems. We place particular 
emphasis on the computation of the pressure of the system which is the effective interaction between the two plates 
modified by the presence of counterions. The effect of asymmetric charge distribution on the plates is analyzed in 
detail. This system, while being somewhat idealized, is a simple model of an array of charged smectic bilayers 
sandwiched between two fixed parallel charged plates. If the charge on the bilayers is uniform then the use of the 
one-dimensional Coulomb interaction is justified.

Another reason for studying this simple system is to check the validity of the various approximation schemes alluded 
to above. The mean-field and strong-coupling approximations in the planar geometry are independent of the 
dimensionality of the problem. This is due simply to the fact the both Poisson-Boltzmann theory as well as the 
strong-coupling theory are one-dimensional effective theories since the corresponding fields depend only on the 
coordinate parallel to the bounding surfaces normals. However, fluctuations about the mean field, or the 
Poisson-Boltmzann configuration, do depend on the dimensionality of the problem as they correspond to thermal Casimir 
or zero-frequency van-der-Waals forces. This makes these model calculations particularly appealing since they can 
clarify the role the fluctuations play in Coulomb systems.

The paper is organized as follows: first we use the Edwards-Lenard path integral formulation for the problem adapted 
to apply to a finite electroneutral system with surface charges. The exact expression obtained for the force between 
the two plates is evaluated numerically and compared with the results of Monte-Carlo simulation for a wide range of 
cases: symmetric/asymmetric surface charges, number of counterions, etc. We find excellent agreement between theory 
and simulation in all cases as we should expect since the theoretical solution is exact.  In the next section we 
compare the results with the Poisson-Boltzmann mean-field approximation and investigate its domain of validity. We 
present two methods for calculating the effects of field fluctuations about the mean-field solution and compare with 
our exact results. The strong-coupling approximation is then applied to the system and a similar comparison with the 
exact results is made.

\section{The counterion model}

We consider a system of charged particles interacting {\em via} a Coulomb potential in one dimension. It can also 
be considered as a system of uniformly charged infinite two dimensional sheets in three dimensions, where the sheets are perpendicular to the direction $x$, are uniformly charged and can only move in this direction. The Hamiltonian for this system is given by
\begin{equation}
\beta {\mathcal H} = -{\beta e^2\over 4}\sum_{\alpha,\beta} z_\alpha z_\beta \vert x_{\alpha}-x_{\beta}\vert,
\label{eq:Hamilt}
\end{equation}
where, $e$ is the unit electric charge, $\beta = 1/k_BT$, $z_\alpha$ is the valence of particle (or sheet) $\alpha$ 
and $x_\alpha$ is its position. We will consider systems in a canonical formulation which are 
overall charge neutral, so that
\begin{equation}
\sum_{\alpha} z_\alpha = 0.
\end{equation}
It is convenient to rewrite the Hamiltonian as
\begin{eqnarray}
\beta {\mathcal H} &=& -{\beta e^2\over 4}\sum_{\alpha,\beta} z_\alpha z_\beta \big[x_\alpha +x_\beta - 2\min(x_\alpha,x_\beta)\big]
\nonumber \\
&=& {\beta e^2\over 2}\sum_{\alpha,\beta} z_\alpha z_\beta  \min(x_\alpha,x_\beta),
\end{eqnarray}
where we have used charge neutrality to go from the first to second equation above. We may now 
write the Hamiltonian as an expectation value over a ``Brownian motion"
\begin{equation}
\beta {\mathcal H} = {1\over 2}\bigg\langle \left[ \sum_{\alpha} z_{\alpha} \psi(x_{\alpha}) \right]^2\bigg\rangle,
\end{equation}
where $\psi$ is Brownian motion with correlation function
\begin{equation}
\langle \psi(x) \psi(y)\rangle = \beta e^2\min(x,y), \label{bmphi}
\end{equation}
and started at the value zero at $x=0$, so
\begin{equation}
\psi(0) = 0.
\end{equation}
The Boltzmann weight for any configuration can thus be written as
\begin{equation}
\exp(-\beta {\mathcal H}) = \langle \exp[i\sum_{\alpha} z_{\alpha} \psi(x_{\alpha})] \rangle.
\end{equation}
In path-integral notation the measure on the Brownian motion $\psi$ as defined by Eq. ({\ref{bmphi}) is given by 
\begin{eqnarray}
 \langle {\cal O}[\psi] \rangle =  \int_{-\infty}^{\infty} d\psi(L) \int_{\psi(0)}^{\psi(L)}
d[\psi] \exp\left( -{1\over 2 e^2\beta}\int_0^L dx\ {\left(d \psi(x) \over dx\right)}^2\right){\cal O}[\psi],\nonumber \\ \label{bmeasure}
\end{eqnarray}
where $L$ is the overall length of the system. 

The sum over $\alpha$ has three distinct contributions: that from the counterions with $1\leq \alpha\leq N$, 
for which from now on we assume have valence $z_\alpha = 1$, and those from the surface charge at $x_0 = 0$, 
with valence $z_0 = -(N+P)/2 $, and from the surface charge at $x_{N+1}=L$, with valence $z_{N+1}= 
-(N-P)/2$. The system is globally electroneutral, consisting of two surface charges and their 
counterions. For the moment, we will assume that $N \pm P$ are even integers, which will be the case if the 
counterions originate by being released into solution from initially neutral surfaces; {\em i.e.}, each surface 
charge is an integer multiple of the unit of charge of the counterions. The generalization of the approach 
due to relaxing the latter assumption is given in the next section. We then have
\begin{equation}
 \sum_{\alpha} z_{\alpha} \psi(x_{\alpha}) = \sum_{i=1}^{N} \psi(x_i) -{N+P\over 2}\psi(0) - {N-P\over 2} \psi(L),
\end{equation}
and the Boltzmann weight for a given configuration is thus
\begin{eqnarray}
BW(\left\{x_i\right\}) =\bigg \langle \exp\left( i\sum_{i=1}^N \psi(x_i) - i\psi(0) {N+P\over 2} - i\psi(L) {N-P\over 2} \right)\bigg\rangle.
\end{eqnarray}
The partition function is then obtained as
\begin{eqnarray}
Z_N &=&\bigg\langle  {1\over N!}\int_0^L \prod_{i=1}^N  dx_i BW(\left\{x_i\right\}) \bigg\rangle \nonumber \\
&=& {1\over N!} \bigg\langle \left(\int_0^L dx \exp(i\psi(x))\right)^N
\;\exp\left( - i\psi(0) {N+P\over 2} - i\psi(L) {N-P\over 2} \right)\bigg\rangle.
\end{eqnarray}
We now introduce an arbitrary fugacity $\kappa$ having the dimensions of inverse length and write
\begin{eqnarray}
\kappa^N Z_N &=&\bigg\langle  \exp\left( - i\psi(0) {N+P\over 2} - i\psi(L) {N-P\over 2} \right)  \nonumber \\
&&\qquad \qquad \times \int_0^{2\pi} {d\lambda\over 2\pi}\sum_{N'=0}^\infty  \exp\left (-i\lambda(N-N')\right)   
{\kappa^{N'}\over N'!} \left(\int_0^L dx \exp(i\psi(x))\right)^{N'}\bigg\rangle \nonumber  \\
 &=& \bigg\langle \int_0^{2\pi} {d\lambda\over 2\pi} 
\exp\left( - i\psi(0) {N+P\over 2} - i\psi(L) {N-P\over 2} -i\lambda N\right)
\exp\left( \kappa \int_0^L dx \exp(i\lambda  +  i \psi(x))\right)\bigg\rangle,
\end{eqnarray}
although, clearly, the value of $\kappa$ should not affect the final physical results since charge 
neutrality will enforce the constraint $N'=N$. This is ensured by the integral over $d\lambda$. If we define 
a shifted Brownian motion, \begin{equation} \phi(x) = \lambda + \psi(x), \end{equation} the starting 
position of the field $\phi$ at $x=0$ is simply $\phi(0) =\lambda$ but the integration over the end point 
remains free and we have
\begin{equation}
Z_N =   \bigg\langle \int_0^{2\pi} {d\phi(0)\over 2\pi} \exp\left( - i\phi(0){N+P\over 2} -i\phi(L){N-P\over 2}\right) 
 \exp\left( \kappa \int_0^L dx \exp( i \phi(x))\right)\bigg\rangle.
\end{equation}  
The electroneutrality constraint has thus been absorbed into the integration over the starting point
of the shifted field $\phi$. 

It is convenient, for what comes later, to generalize the result to the case where the counterions have 
valence $z_\alpha = q,~1 \le \alpha \le N$. This is easily accomplished by replacing $e$ by $qe$ in the 
measure for the Brownian motion in Eq. (\ref{bmeasure}). This follows from the forgoing analysis and 
rescaling the field $\phi(x) \to \phi(x)/q$. In path integral notation we then have
\begin{eqnarray}
Z_N &=&    \int_0^{2\pi} {d\phi(0)\over 2\pi}\int_{-\infty}^{\infty}d\phi(L)   
\exp\left( - i\phi(0){N+P\over 2} -i\phi(L){N-P\over 2}\right)\nonumber\\
 &&\qquad \qquad \times \int_{\phi(0)}^{\phi(L)} d[\phi]
\exp\left(-\int_0^Ldx\  \left[{1\over 2 q^2e^2\beta}{\left(d\phi(x)\over dx\right)}^2 -\kappa\, \exp\left(i\phi(x)\right)  \right]\right)
\label{eq:ZN_FT}
\end{eqnarray}
as our final functional integral expression for the partition function.

\section{Exact evaluation of the partition function}
\label{sec:exact}

We use the Feynman formula, which is the Euclidean version of the equivalence between 
path integrals and the propagator for the Schr\"odinger equation. We have that
\begin{eqnarray} 
&&\int_{-\infty}^{\infty} d\phi(L)\int_{z}^{\phi(L)} d[\phi] \exp\left(-\int_0^Ldx\  
\left[{1\over 2 q^2e^2\beta}{\left(d\phi(x)\over dx\right)}^2 -\kappa\, \exp\left(i\phi(x)\right)  \right]\right)  \exp\left( - i\phi(L){N-P\over 2}\right) \nonumber \\
&=& \exp(-L\hat H)  \exp\left( - iz{N-P\over 2}\right),
\end{eqnarray} 
where $q$ is the counterion valence and where $\hat H$ is the complex operator/Hamiltonian given by
\begin{equation}
\hat H =- {q^2e^2 \beta\over 2} {d^2\over dz^2} - \kappa\, \exp(iz)\;.
\end{equation}
The partition function can thus be expressed as  
\begin{equation}
\kappa^N Z_N(L) = \int_0^{2\pi} {dz\over 2\pi} \exp\left( - i z{N+P\over 2}\right) \exp(-L\hat H)  
\ \exp\left( - iz{N-P\over 2}\right). \label{cons}
\end{equation}

We should note that the operator $\hat H$ is complex in our problem but in the case of symmetric electrolytes 
\cite{ed62} it is real. Because of this we analyse the problem via Fourier analysis
rather than on the basis of the eigenfunctions of the operator $\hat H$.

We define integers $M_1 = (N+P)/2$ and $M_2 = (N-P)/2$ and the surface charges are $-M_1qe$ and $-M_2qe$ at 
$x=0$ and $x=L$, respectively, with $M_1+M_2=N$. A quick preliminary check of our approach is to consider 
the perfect gas limit where $e$ = 0. In this case we find trivially that
\begin{equation}
\kappa^N Z_N(L) = \kappa^N {L^{M_1+M_2}\over (M_1+M_2) !}\;,  
\end{equation}
and thus
\begin{equation}
Z_N(L) = {L^N\over N !} \;, 
\end{equation}
which is the perfect gas result.  In order to proceed further we consider the evaluation of
\begin{equation} 
f(z;L) = \exp(-L\hat H) f(z;0).
\end{equation}
We can write \cite{de98}
\begin{equation}
f(z) = \sum_n a(n,L) \exp(in z)\;,
\end{equation}
and the induced evolution equation for the Fourier coefficients is 
\begin{equation}
{da(n,L)\over dL} = -{n^2 q^2e^2 \beta\over 2} a(n,L) + \kappa\, a(n-1,L)\;.
\end{equation}
In our problem we have $f(z,M_2;0) = \exp\left( - izM_2\right)$ and thus the initial condition on the 
Fourier coefficients $a(n,M_2;0) = \delta_{n,-M_2}$. We define $a_n = \kappa^n b_n$ and find that
\begin{equation}
{db(n,M_2;L)\over dL} = -{n^2 q^2e^2 \beta\over 2} b(n,M_2;L) + b(n-1,M_2;L)\;, \label{beq}
\end{equation}
with the initial condition $b(n,M_2;L) = \kappa^n \delta_{n,-M_2}$ and from Eq. (\ref{cons}) we find that  
we can express the partition function for  system of unit charges $M_1$ and $M_2$  on the left and right boundaries 
respectively as  
\begin{equation}
Z_{M_1,M_2}(L) = b(M_1,M_2;L)\;,
\end{equation}
but with the initial condition $b_n(0,M_2) = \delta_{n,-M_2}$; as we expect, the fugacity $\kappa$, which 
was introduced on dimensional grounds, does not enter in the final physical result. The evolution 
Eq. ({\ref{beq}) can be written as
\begin{equation}
p_{M_1,M_2}  =  {1\over \beta b(M_1,M_2;L)}{db(M_1,M_2;L)\over dL} 
 =   -{M_1^2 q^2e^2\over 2}  + {Z_{M_1-1,M_2}\over  \beta Z_{M_1,M_2}} \;,\label{eqpres}
\end{equation}
where $p_{M_1,M_2}$ is the pressure of the system. Clearly, the last term makes a positive contribution to the 
pressure as it is the ratio of two  partition functions for systems with different surface charges.  
Equation (\ref{eqpres}) has a nice physical  interpretation. The
second term is, up to the factor of $\beta$, the average density of counterions at $z=0$ because 
it is the  restricted partition function where at least one particle is on the surface $z=0$ (thus giving a
reduction in the surface charge to $M-1$) normalized by the full partition function. We can thus write
\begin{equation}
p_{M_1,M_2} = -{\sigma_1^2\over 2} + {1\over \beta}\rho(0)\;,
\label{eq:contact_theorem}
\end{equation}
where $\rho(x) = \langle\sum_{i=1}^N \delta(x-x_i)\rangle $ is the average value of the counterion density 
at the surface $z=0$ and $\sigma_1 = -M_1qe$ the corresponding surface charge. However, this is simply the contact-value theorem for electrostatic systems known to be exact in any dimension 
\cite{contact0,contact1,contact2,ver90}. In fact the contact-value theorem can be demonstrated {\em via} an 
extension of the path integral methods used here to higher dimensional systems with (parallel) planar 
geometries \cite{de03}. The average counterion density at the point $x$ can be shown \cite{ed62,de98} to be equal to
\begin{eqnarray}
\rho(x) &=& \langle \exp(i\phi(x))\rangle \nonumber \\
&=& {1\over Z_{M_1,M_2}}\int_0^{2\pi} {dz\over 2\pi} \exp\left( - i M_1z\right) 
\exp(-x\hat H)\exp(iz) \exp(-(L-x)\hat H)\exp\left( - i M_2z\right). \nonumber
\end{eqnarray}
If we set $x=0$ in this formula we obtain 
\begin{equation}
\rho(0) =  {Z_{M_1-1,M_2}\over   Z_{M_1,M_2}}\;, 
\end{equation}
in agreement with the previous discussion.

\subsection{The general formalism}

The charges on the surfaces at $z=0$ and $L$ can be denoted $\sigma_1$ and $\sigma_2$
respectively. Without loss of generality we may assume that $q>0$ and $|\sigma_2|\geq |\sigma_1|$, which
allows $\sigma_1$ and $\sigma_2$ to be of opposite sign, and we can define 
the charge asymmetry parameter $\zeta = \sigma_2/\sigma_1$ with $-1 \le \zeta \le 1$.
All other cases can be mapped onto this interval with appropriate rescaling of the parameters. 
Note that the values $\zeta = 1$ and $\zeta = -1$ represent special cases of  symmetric system 
with $\sigma_1=\sigma_2$,  and antisymmetric system with $\sigma_1=-\sigma_2$ and no counterions 
between surfaces (which thus reduces to the trivial case of a planar capacitor). 

For the above analysis we see that
\begin{equation}
\zeta = \frac{N-P}{N+P}\;,
\end{equation}
$\sigma_1 = -M_1qe$ and $\sigma_2 = -M_2qe$. The independent parameters are conveniently chosen as 
$\sigma_1,\zeta,N$. With $M_1$ and $M_2$ integers, the analysis so far allows only for particular discrete 
values of $\zeta$. However, the approach can be extended to accommodate all values of $\zeta$ and we 
state the generalization of the algorithm here. Charge neutrality determines that the valence $q$ of the 
counterion satisfies

\begin{equation}
q = -\frac{\sigma_1+\sigma_2}{Ne} = -\frac{\sigma_1}{Ne}(1+\zeta)\;.
\end{equation}
We are allowing that $q$ need not be an integer.

In addition to the Fourier coefficients $b(n,M_2;L)$ we introduce Fourier coefficients
$c(n,M_1;L)$. Equation (\ref{beq}) gives the evolution of the $b(n,M_2;L)$ from $z=0$, the 
surface with assigned charge $\sigma_1$. The coefficients $c(n,M_1;L)$ obey a similar evolution
equation but now with the surface charge assignment reversed; the charge at $z=0$ is now $\sigma_2$.
These two sets of coefficients correspond to complementary approaches; the set $\{b\}$ describes
evolution from the surface with charge $\sigma_1$  to that with charge $\sigma_2$, and vice-versa 
for the set $\{c\}$.

We define
\begin{equation}
\alpha = \frac{1}{1+\zeta}\;,~~~M_1 = \mbox{Int}(\alpha N)\;,~~~\eta_1 = \alpha N - M_1\;,~~~
\eta_2 = 1 - \eta_1\;,~~~M_2 = N - M_1 -1\;.
\end{equation}
The evolution equations now become
\begin{eqnarray}
\frac{db(n,M_2;L)}{dL}&=&\frac{(n-\eta_2)^2}{2}\beta q^2e^2 b(n,M_2;L) + b(n-1,M_2;L)\;, \label{beq_gen} \\
\frac{dc(n,M_1;L)}{dL}&=&\frac{(n-\eta_1)^2}{2}\beta q^2e^2 c(n,M_1;L) + c(n-1,M_1;L)\;. \label{ceq_gen}
\end{eqnarray}
The initial conditions are $b(n,M_2;L=0) = \delta_{n,-M_2}$,~$c(n,M_1;L=0)=\delta_{n,-M_1}$.

Then the partition function is given by alternative formulas $Z(\sigma_1,\zeta,N;L) = b(M_1+1,M_2;L) = c(M_2+1,M_1;L)$,
and the pressure is given equivalently by
\begin{equation}
p(L) = \left\{\begin{array}{l}
\displaystyle -\frac{(M_1+1-\eta_2)^2}{2}q^2e^2~+~\frac{b(M_1,M_2;L)}{Z(\sigma_1,\zeta,N;L)} \\
\\
\displaystyle -\frac{(M_2+1-\eta_1)^2}{2}q^2e^2~+~\frac{c(M_2,M_1;L)}{Z(\sigma_1,\zeta,N;L)}\;. 
\end{array}
\right.
\label{eq:P_exact}
\end{equation}

The partition function $Z(\sigma_1,\zeta,N;L)$ is also given by
\begin{equation}
Z(\sigma_1,\zeta,N;L) = \sum_{j=-M_1}^{M_2+1}\,b(-j+1,M_2;L-x)\,c(j,M_1;x),
\end{equation}
for any $x,\; 0 \le x \le L$. This acts as a check on the numerics.

The counterion number density $\rho(x; \sigma_1,\zeta,N,L)$ is given by
\begin{equation}
\rho(x;\sigma_1,\zeta,N,L) = \frac{1}{Z(\sigma_1,\zeta,N;L)}\sum_{j=-M_1}^{M_2+1}\,b(-j,M_2;L-x)\,c(j,M_1;x),
\end{equation}
and another check on the numerics is that
\begin{equation}
\int dx \rho(x; \sigma_1,\zeta,N,L)~=~N,
\end{equation}
{\em i.e.} the system is electroneutral.

\section{Approximate evaluations of the partition function}

A traditional approach to the study of charged (bio)colloidal systems is the mean-field 
Poisson-Boltzmann (PB) formalism which is applicable for weak surface charges, low counter-ion 
valency and high temperature \cite{ve48}. The limitations of this approach are evident
when applied to highly-charged systems where counterion-mediated interactions 
between charged bodies start to deviate substantially from the accepted mean-field wisdom 
\cite{hoda,Naji}. One of the recent fundamental advances in this field has been the 
systematization of these non-PB effects based on the notions of {\em weak} and {\em strong} 
coupling approximations. The latter approach has been pioneered by Rouzina and Bloomfield 
\cite{ro96}, elaborated later by Shklovskii {\em et al.} \cite{gr02}, Levin {\em et al.} 
\cite{le02}, and brought into final form by Netz {\em et al.} \cite{ne01,mo02,hoda,Naji}. 
These two approximations allow for an explicit and exact treatment of charged systems at 
two disjoint {\em limiting conditions} whereas the parameter space in-between can be 
analyzed only approximately \cite{sa06,bu04,ne01,mo02,ch06,ro06} and is mostly accessible 
solely {\em via} computer simulations 
\cite{hoda,Naji,ne01,mo02,ch06,ro06,gu84,br86,va91,kj92,jh07,tr06}.

Both the weak- and the strong-coupling approximations are based on a functional integral or 
field-theoretic representation \cite{po89,po90} of the grand canonical partition function 
for a system composed of fixed surface charges with intervening mobile counterions, and 
depend on the value of a single dimensionless coupling parameter $\Xi$ \cite{ne01,mo02}. 
In three dimensions, $\Xi$ is proportional to the ratio between two relevant length 
scales, namely, the Bjerrum length and the Gouy-Chapman length. The Bjerrum length, defined as
$\ell_{\mathrm{B}}=e^2/(4\pi\varepsilon\varepsilon_0 k_{\mathrm{B}}T)$, is the distance
at which two unit charges interact with thermal energy $k_{\mathrm{B}}T$  
(in water at room temperature, one has $\ell_{\mathrm{B}}\simeq 0.7$nm). If the charge 
valency of the counterions is $q$ then the corresponding length scales as $q^2 \ell_{\mathrm{B}}$. 
Similarly, the Gouy-Chapman length, defined as $\mu=e/(2\pi q\ell_{\mathrm{B}}|\sigma|)$, is
the distance at which a counterion interacts with a macromolecular surface (of 
surface charge density $\sigma$) with an energy equal to $k_{\mathrm{B}}T$. The 3D electrostatic 
coupling parameter measures the competition between ion-ion and ion-surface interactions
and is given by \cite{ne01,mo02}
\begin{equation}
\Xi = \Xi_{\mathrm{3D}}\equiv q^2 \ell_{\mathrm{B}}/\mu=2\pi q^3 \ell_{\mathrm{B}}^2|\sigma|/e \;. \label{xi3D}
\end{equation}
Physically, the weak-coupling (WC) regime ${\Xi}\ll 1$ (appropriate for low valency 
counterions and/or weakly charged surfaces), is characterized by the fact that the width, $\mu$, of 
the counterion layer near the surfaces is much larger than the separation between two neighboring 
counterions in solution, and thus the counterion layer behaves basically as a 
three-dimensional gas. Each counterion in this case interacts with many others and the 
collective mean-field approach of the Poisson-Boltzmann (PB) type is completely justified. 

On the other hand in the strong-coupling (SC) regime ${\Xi}\gg 1$ (appropriate for high 
valency counterions and/or highly charged surfaces), the mean distance between counterions, 
$a_\bot =| \sigma|/qe$, is much larger than the layer width ({\em i.e.}, $a_\bot/\mu\sim 
\sqrt{\Xi}\gg 1$), indicating that the counterions are highly localized laterally and form 
a strongly correlated quasi-two-dimensional layer next to a charged surface. In this case, 
the weak-coupling approach breaks down due to strong counterion-surface and 
counterion-counterion correlations. Since counterions can move almost independently from 
the others along the direction perpendicular to the surface, the collective many-body 
effects that enable a mean-field description are absent, necessitating a complementary SC 
description \cite{ne01,mo02}.

Formally, the WC limit can be identified with the saddle-point 
approximation of the field theoretic representation of the grand canonical partition 
function, and reduces to the mean-field PB theory at the lowest order for 
$\Xi\rightarrow 0$. The quadratic fluctuations around the mean field provide a second-order 
correction to the mean-field solution for small $\Xi<1$ 
\cite{po89,po90,ne99,ka99,at88,pi98,ha01,la02}. The SC approximation has no PB-like 
collective mean-field \cite{ne01,mo02} since it is formally equivalent to a single particle 
description obtained from a systematic $1/\Xi$ expansion in the limit $\Xi\rightarrow 
\infty$, and corresponds to two lowest order terms in the virial expansion of the grand 
canonical partition function. The consequences and the formalism of these two limits of the 
Coulomb fluid description have been explored widely and in detail (for reviews, see Refs. 
\cite{hoda,Naji}).

The concept of weak- and strong-coupling electrostatics can be easily generalized to other dimensions. 
In one dimension (1D), the  Coulomb interaction between two unit charges may be written as 
$v_{\mathrm{1D}}(x) = -|x|k_BT/\ell_{\mathrm{B}}$, where $\ell_{\mathrm{B}} = 2k_BT/ e^2$
may be regarded as a 1D Bjerrum length. Likewise, the interaction strength with the boundary charges
in an asymmetric system may be characterized by the 1D Gouy-Chapman lengths
$\mu_1 = \ell_{\mathrm{B}} e/q|\sigma_1|$ and  $\mu_2 = \ell_{\mathrm{B}} e/q|\sigma_2|$. We may
proceed by using only $\mu \equiv \mu_1 = \ell_{\mathrm{B}} e/q|\sigma_1|$ in what follows since 
for any given asymmetry parameter $\zeta = \sigma_2/\sigma_1$, we have $\mu_2 = \mu/|\zeta|$.  
The ratio 
\begin{equation}
\Xi_{\mathrm{1D}} \equiv q^2\mu/\ell_{\mathrm{B}}
\label{eq:x1d}
\end{equation}
 is then the corresponding electrostatic 
coupling parameter in 1D. (On a more formal level, this definition for the 
coupling parameter may be established by looking at the field-theoretical representation 
(\ref{eq:ZN_FT}) and rescaling the coordinates as $x_i\rightarrow x_i/\mu_1$.) Note that in 
1D the coupling parameter is related to the asymmetry parameter by virtue of the 
electroneutrality condition
\begin{equation}
  N qe = -(\sigma_1 + \sigma_2)\;,
 \label{eq:neutral}
\end{equation}
where $N$ is the number of counterions of valency $q$, as
\begin{equation}
  \Xi_{\mathrm{1D}} = \frac{\zeta +1}{N}. 
  \label{eq:rescl_neutral}
\end{equation}

Thus, it may be expected that in the limit $N\rightarrow \infty$, contrary to the 3D case, the mean-field theory 
corresponding to this system becomes exact! On the other hand, the SC description may be 
expected to follow simply for $N=1$. These limiting cases will be discussed further in the 
forthcoming sections and will be compared with the exact results and Monte-Carlo 
simulations. One should bear in mind that the 1D system considered here corresponds 
to a 3D system of mobile charged plates (membranes) confined between two fixed planar 
charged walls and thus, such a 1D system with even a single ``counterion" would have a 
completely meaningful thermodynamic behavior. 

We note also that both the PB theory ($\Xi\rightarrow 0$) and the SC theory ($\Xi\rightarrow \infty$) 
for uniformly charged plates are one-dimensional theories and should remain valid in 3D as well as in 
the 1D case that we are studying here.  However, even though the mean-field solution depends only 
on the transverse coordinate, the field fluctuations about the mean-field solution will depend on all 
coordinates and so it is clear that fluctuations corrections to the mean-field contribution depend 
on the dimensionality of the system: they will be different for a 1D system than for a 3D system.

\subsection{\label{sec:weak_coupling}Weak-coupling limit: Poisson-Boltzmann theory}

The expression for the partition function $Z_N$ in Eq. (\ref{eq:ZN_FT})  is up to multiplicative 
constants given by the functional integral
\begin{equation}
Z_N = \int d[\phi] \exp\left(-S[\phi]\right)
\end{equation}
where the action $S$ is given by
\begin{equation}
S[\phi] = \int_0^L dx\left[ {1\over 2q^2 e^2 \beta} \left(d\phi\over dx\right)^2 - \kappa\exp(i\phi)
-i\phi{\sigma_1\over qe}\delta(z) -i\phi{\sigma_2\over qe}\delta(z-L)\right]
\label{eq:ZN_FT2}
\end{equation}
In the weak-coupling regime, the leading contribution to the partition function comes from 
the saddle-point configuration, $\phi_0(x)$, of the action in Eq. (\ref{eq:ZN_FT2})  
\cite{po89,po90}, where the field $\phi_0(x) = i\psi_0(x)$ turns out to be imaginary and is proportional to the mean-field electrostatic potential. The saddle-point configuration can be 
straightforwardly translated into a solution of the PB equation and corresponds to an exact asymptotic 
result in the limit $\Xi\rightarrow 0$ \cite{ne01,mo02,ne99}. The PB equation for the potential, 
$\psi_0(z)$, can be written as 
\cite{intro2,ve48}
\begin{equation}
\frac{\mathrm{d}^2 \psi_0(x)}{\mathrm{d}x^2}=-q^2e^2\beta\kappa\, e^{-\psi_0(x)},
\label{eq:PBeq}
\end{equation}
with boundary conditions  
\begin{eqnarray}
&&\left.\frac{\mathrm{d}\psi_0}{\mathrm{d}x}\right\vert_{0}=-{\sigma_1\beta q e}= \frac{2}{\mu},\nonumber\\
&&\left.\frac{\mathrm{d}\psi_0}{\mathrm{d}x}\right\vert_{L\phantom{-}}= {\sigma_2\beta qe}=-\frac{2\zeta}{\mu}
\label{BC-rep}
\end{eqnarray}
stemming from the electroneutrality of the system. 

Integration of the PB equation gives rise to the first integral of the system of the form
\begin{equation}
\beta p_0 = -\frac{1}{2q^2e^2\beta}\bigg(\frac{\mathrm{d}\psi_0}{\mathrm{d}x}\bigg)^2 + \rho_0(x),
\label{rep-pressure}
\end{equation}
where the constant $p_0$ is nothing but the mean-field PB pressure acting between the 
bounding surfaces \cite{intro2} and
\begin{equation}
 \rho_0(x) = \kappa\,e^{-\psi_0(x)},
\end{equation}
is the PB number density profile of counterions between the surfaces. As remarked earlier in section 
\ref{sec:exact}, the choice for $\kappa$ is arbitrary since it corresponds to a choice of origin for the 
potential $\psi(x)$. In section \ref{sec:exact} we couched the Fourier solution in terms of Fourier 
coefficients $a_n = \kappa^nb_n$ with the choice $\kappa=1$, and so we adopt this choice here, too.

The nature of the solution $\psi_0(x)$ obviously crucially depends on the sign of the pressure $p_0$ 
\cite{la99,be07}. Different forms are obtained for positive and negative pressures, corresponding to 
repulsion and attraction between the bounding surfaces respectively and were derived previously \cite{ka08}.

Guided by the contact-value theorem Eq. (\ref{eq:contact_theorem}), which we emphasize is 
an exact result, we express the Poisson-Boltzmann, or mean-field, pressure in rescaled units. The repulsive 
PB pressure is given by

\begin{equation}
\tilde p_0=\frac{2p_0}{\sigma_1^2}=\frac{\beta p_0\ell_{\mathrm{B}}}{(\sigma_1/e)^2} = \tilde \alpha^2. 
\label{pres-pb-rep}
\end{equation}
and all the other quantities have been defined above.
Furthermore,  $\tilde \alpha = \alpha \mu$ and its value is given by the solution of 
\begin{equation}
\tan\,(\alpha L)=\frac{\alpha(\zeta+1)\mu}{\alpha^2\mu^2-\zeta}.
\end{equation}

For $\zeta < 0$, and only then, can the PB pressure be attractive. The attractive PB pressure is given by
\begin{equation}
\tilde p_0=\frac{2p_0}{\sigma_1^2}=\frac{\beta p_0\ell_{\mathrm{B}}}{(\sigma_1/e)^2} = -\tilde \alpha^2. 
\label{pres-pb-att}
\end{equation}
where $\tilde \alpha$ is now given as a solution of 
\begin{equation}
{\rm coth}\,(\alpha L)=-\frac{\zeta+\mu^2\alpha^2}{\mu\alpha(1+\zeta)}.
\end{equation}

The limiting case of zero pressure can be obtained straightforwardly from either limit. The interaction 
pressure can be obviously computed for any value of the asymmetry parameter, but note that any value of 
$\zeta$ can be mapped onto the interval $-1 < \zeta < 1$ and the pressure only need to be evaluated in that 
interval of $\zeta$ values. Note again that the PB interaction pressure is the same for a 1D as well as for 
a 3D system.

\subsection{Weak-coupling limit: fluctuations}

Fluctuations around the mean field depend on the dimensionality of the system and are different for a 1D 
than for a 3D system. In order to proceed, one needs to evaluate the appropriate Hessian of the field action 
in the partition function and study its fluctuation spectrum (see Refs. \cite{po89,po90,ka07} for more 
details). The Hessian of the field about the mean-field value is given by
\begin{equation}
{\delta^2 S\over \delta \phi(x)\phi(y)}\vert_{\phi=i\psi_0}= H(x,x') ={1\over \beta e^2q^2}
\left[ -{d^2\over dx^2} + \beta e^2 q^2 \rho_0(x)\right]\delta(x-x'),
\end{equation}
$\rho_0(x) =\exp(-\psi_0(x))$ is the mean-field  PB counterion density. Hence,
\begin{equation}
\beta e^2 q^2 \rho_0(x)= \left\{
	\begin{array}{ll}
	 \cfrac{2\alpha^2}{\cos^2\alpha (x-x_0)} & \quad p_0 > 0,\\
	\\
	 \cfrac{2\alpha^2}{\sinh^2\alpha (x-x_0)} & \quad p_0 < 0.
	\end{array}
\right.
\label{sign-eigen}
\end{equation}
The corresponding correction, ${\mathcal F}_2$,  to the free energy of the system
comes from  the functional integral
\begin{equation}
{\cal A}(L) = \int dx dy \int_{\psi(0) = x}^{\psi(L )=y } d[\psi] \exp(-s[\psi]),\label{fluc}
\end{equation} 
where
\begin{equation}
s[\psi] = {1\over 2}\int\!\!\!\int {\psi(x)}H(x,x') {\psi(x')} dx dx',
\end{equation} 
so that the corresponding fluctuation part of the free energy can be obtained in the form
\begin{equation}
  \beta {\mathcal F}_2 = - \ln\, {{\cal A}(L)}.
\label{trace1}
\end{equation}

The functional integral Eq. (\ref{fluc}) can be evaluated exactly, in two different ways. The first 
method is based on the use of the argument principle \cite{at88} converting the discrete sum of eigenvalues 
of the Hessian operator into the logarithm of the secular determinant ${\mathcal D}_\lambda$ of the same 
operator. The second method is based on the Pauli-van Vleck approach to calculating the functional integral 
of a general harmonic kernel. Both results are the same.

In the first approach the trace-log of the Hessian can be written equivalently in the form that was derived 
for a 3D case \cite{po89,po90,at88} but can be used in a trivially modified form also for the 1D case under 
consideration. It gives
\begin{equation}
  \beta {\mathcal F}_2 = {\textstyle\frac{1}{2}}{\rm Tr ~ln}\,H(x,x')
= {\textstyle\frac{1}{2}}\ln\frac{{\mathcal D}_1}{{\mathcal D}_0},
\label{trace}
\end{equation}
where ${\mathcal D}_{\lambda}$ is the secular determinant, {\em i.e.} the determinant of the coefficients 
corresponding to the appropriate boundary condition in the solution of the eigenvalue equation, that can be 
derived from solutions of
\begin{equation}
\Bigl(\frac{
d^2}{
d x^2} -\lambda \beta e^2q^2\rho_0(x)\Bigr) f_{\lambda}(x)=0.
\label{eigF}
\end{equation}
One can now  write ${\mathcal A}(L)$ for the quotient  $({{\mathcal D}_1}/{{\mathcal D}_0})^{-1/2}$, since the 
secular determinant depends explicitly on the value of the inter-surface spacing, $L$.   Using the fact 
\cite{ka07,po89,po90} that the two linearly independent solutions of Eq. (\ref{eigF}) for $\lambda = 1$ 
in the repulsive regime are
\begin{equation}
f^{(1)}(x) = \tan{\alpha x} \qquad {\rm and} \qquad f^{(2)}(x) = 1 + \alpha x\tan{\alpha x}
\end{equation}
the secular determinant for the symmetric case $\zeta =1$ comes out as
\begin{equation}
{\cal D}_1(L) =  \alpha \sec^2\left(\alpha {L\over 2}\right) \left[\tan\left(\alpha {L\over 2}\right) + 
\alpha {L\over 2} \sec^2\left(\alpha {L\over 2}\right)\right],
\end{equation}
and the corresponding free energy contribution from the quadratic fluctuations is thus given by
\begin{equation}
\beta {\mathcal F}_2= \frac{1}{2}\, \ln\left\{\alpha \sec^2\left(\alpha {L\over 2}\right) 
\left[\tan\left(\alpha {L\over 2}\right) + \alpha {L\over 2} \sec^2\left(\alpha {L\over 2}\right)\right]\right\}. 
\label{second-f}
\end{equation}

The fluctuation free energy can be  regularized so that 
all irrelevant constants, {\em i.e.} all the terms not depending on the separation between the bounding 
surfaces, are dropped, amounting to a rescaling ${\mathcal F}_2(L) \rightarrow {\mathcal F}_2(L) 
- {\mathcal F}_2(L \rightarrow \infty)$. This corresponds to a subtraction of the part of the free 
energy for two separate boundaries at infinite separation from the total free energy.

In the second method we compute the fluctuations about the mean-field path using the Pauli-van Vleck 
approach \cite{kle06,deho05,deho07}. As before the thickness of the film is $L$ we take the leftmost and rightmost points of the film 
to be at $x=0$ and $x=L$ respectively. Clearly the action of the classical path minimizing $s$ 
in Eq. (\ref{fluc}) is a quadratic function of the initial and final points of the fluctuating 
field $\psi$. The generalized Pauli-van Vleck formula tells us that
\begin{equation}  
\int_{\psi(0) = x}^{\psi(L)=y } d[\psi] \exp(-s[\psi]) = \left( -{\frac{1}{2\pi}{\partial s_c[x,y]\over \partial
x\partial y}}\right)^{1\over 2} \exp\left(-s_c[x,y]\right),
\end{equation}
where $s_c$ is the classical action minimizing $s$. As the action is quadratic we may write
\begin{equation}
 s_c[x,y] = {1\over 2}\left[ f(L) x^2 +g(L) y^2 -2 h(L) x y\right].
\end{equation}
In the case of  a  symmetric charge distribution we have $f = g$ and thus the fluctuation term is given by
\begin{equation}
{\cal A}(L) = \left( {2\pi h(L)\over f^2(L) -h^2(L)}\right)^{1\over 2}.
\end{equation}
Solving the equations of motion we find that
\begin{eqnarray}
\beta q^2 e^2f(L) &= & \beta q^2 e^2 g(L) = {\alpha\sec^2(\alpha {L / 2})\over \ 2\tan(\alpha {L / 2})} + 
{\alpha [\tan(\alpha {L / 2}) + (\alpha {L / 2}) \sec^2({\alpha {L / 2}})] \over 2[ 1+ (\alpha {L / 2}) 
\tan(\alpha {L / 2})]}, \nonumber\\  
\beta q^2 e^2h(L)& =& {\alpha \over 2\tan(\alpha {L / 2}) [1+ (\alpha {L / 2}) \tan(\alpha {L / 2})]}.
\end{eqnarray}
Putting all this together then yields
\begin{equation}
{\cal A}(L)  = \left({\pi \over  \alpha \sec^2(\alpha {L / 2}) 
[\tan(\alpha {L / 2}) + (\alpha {L / 2}) \sec^2(\alpha {L / 2})]}\right)^{1\over 2}
\end{equation}
and the contribution to the free energy due to the fluctuations about mean-field free energy is 
again obtained, up to irrelevant constants disappearing upon regularization, as in Eq. ({\ref{second-f}).

As we already stated the fluctuation part of the interaction free energy for 3D and 1D are completely 
different (due to fluctuations in the plane of the film in 3D which are not present in 1D), though 
the mean-field Poisson-Boltzmann result is exactly the same.
In particular, the scaling of the fluctuations contribution to the 
interaction pressure, say $p_2$, with the inter-surface distance $L$ turns out to be very different.
In 3D, one finds that  \cite{ka08}
\begin{equation}
   p_2(L) \sim - ~\Xi_{\mathrm{3D}} ~\frac{\ln L}{L^3}, 
\end{equation}
whereas in 1D we find  
\begin{equation}
   p_2(L) \sim -~\Xi_{\mathrm{1D}} ~\frac{1}{L}
\end{equation}
for same-sign surfaces ($\zeta\geq 0$) at sufficiently large separations. Note that the coupling parameter $\Xi$ is 
defined differently in 1D, Eq. (\ref{eq:x1d}), and 3D, Eq. (\ref{xi3D}). As one goes to the weak-coupling limit in 1D 
for $N \rightarrow \infty$ the fluctuation term obviously makes a vanishing contribution to the total pressure and is 
thus not particularly important. The logarithmic dependence in 3D is a direct consequence of the contribution from the 
in-plane modes. In this case, the mean-field pressure scales with the inverse square of the separation as $p_0(L) \sim 
1/L^2$ irrespective of the dimensionality of the system.

Note that the fluctuations part is always attractive, reflecting the fact that electrostatic correlations mediated by 
counterions always favor attraction between the charged boundaries. Within the WC analysis the fluctuations are always 
assumed to be small as compared with the leading order mean-field contribution. Thus, the total pressure, $p_0+p_2$, 
is dominated by the mean-field contribution (except at the equilibrium point where $p_0=0$ \cite{ka08}) and in 
particular for same-sign surface charges, it still remains repulsive.

It is also interesting to note that the magnitude of fluctuation pressure relative to the mean-field
pressure decreases with the separation distance $L$ in 3D, while in 1D, it increases with $L$. 
For same-sign surfaces, the ratio $|p_2/p_0|$ scales as $\sim \ln L/L$ in 3D, while it scales 
as $\sim L$ in 1D. This indicates qualitatively different distance-dependent behaviors for the fluctuations
and thus, qualitatively different regimes of validity for the loop-expansion approach in 1D and 3D. 
The latter is determined by assuming that $|p_2/p_0|\ll 1$. Hence, the weak-coupling validity 
criterion for same-sign surfaces in 3D reads 
\begin{equation}
\Xi_{\mathrm{3D}} < \frac{L}{\ln L}. 
\end{equation}
This means that at a given coupling parameter, the WC analysis becomes increasingly more accurate at larger 
separations, while as the surfaces get closer a smaller coupling parameter needs to be chosen. In 1D, one has
\begin{equation}
\Xi_{\mathrm{1D}} < \frac{1}{L}, 
\end{equation}
which indicates the opposite trend. In particular, for given $\zeta$ and noting from Eq. (\ref{eq:rescl_neutral}) 
that $\Xi_{\mathrm{1D}} = (1+\zeta)/N$, the WC analysis becomes increasingly more accurate as the separation decreases for 
fixed $N$, or as $N$ increases for fixed separation.

\subsection{Strong-coupling limit}

The strong-coupling approximation coincides with the lowest order of a non-trivial expansion of the partition 
function in terms of the fugacities of the counterions. This expansion may be expressed as a $1/\Xi$ series 
expansion \cite{ne01,mo02}, whose leading order term ($\Xi\rightarrow \infty$) corresponds to the SC theory. 
We will not delve into the strong-coupling expansion in more detail since it has been exhaustively reviewed 
in the literature \cite{hoda,Naji,ne01,mo02}.

At leading order, the SC free energy is obtained as
\begin{equation}
{\mathcal F}=W_0-Nk_\mathrm{B}T\,\ln\int e^{-\beta(W_1+W_2)}\mathrm{d}V,
\label{defF}
\end{equation}
where $W_0$ is electrostatic interaction energy of charged surfaces
\begin{equation}
W_0=-\frac{\sigma_1\sigma_2}{2}\,L,
\end{equation}
with $S$ representing surface area of each plate, and $W_1$ and $W_2$ are electrostatic interaction
energies between a single counterion and individual charged surfaces, {\em i.e.}
\begin{equation}
W_1=-\frac{qe\sigma_1}{2}x,
\qquad
W_2=-\frac{qe\sigma_2}{2}\big(L-x\big).
\end{equation}

Since in the strong-coupling regime the free energy is given {\em via} simple quadratures, it is much 
simpler to evaluate it than at the weak-coupling level. Defining the rescaled free energy

\begin{equation}
\tilde{\mathcal F}=\frac{2}{\sigma_1^2\mu}{\mathcal F},
\end{equation}
we obtain
\begin{equation}
\frac{\tilde {\mathcal F}}{\tilde S}=(1+\zeta^2)\frac{\tilde L}{2} 
-(1+\zeta)\,\ln\,\sinh\bigg[(1-\zeta)\frac{\tilde L}{2}\bigg],
\end{equation}
where ${\tilde L} = L/\mu$. Differentiating the free energy with respect to the surface-surface distance 
$\tilde L $ we get the corresponding pressure acting between the bounding surfaces
\begin{equation}
\tilde p(\tilde L)=-\frac{1}{2}(1+\zeta^2)+\frac{1}{2}(1-\zeta^2)\,\coth\bigg[(1-\zeta)\frac{\tilde L}{2}\bigg],
\label{eq:P_SC}
\end{equation}
where, following the discussion in section \ref{sec:weak_coupling}, we have defined 
$\tilde p(\tilde L) = 2p(L)/\sigma_1^2$. Note that the SC pressure can become attractive for {\em both} 
like-charged and oppositely charged surfaces which contrasts with the mean-field theory that does not allow 
attraction between like-charged surfaces. This is because of the strong electrostatic correlations mediated 
by counterions between the charged surfaces for $\Xi\gg 1$ and has been investigated thoroughly before for 
equally charged surfaces in \cite{ne01,mo02} and for asymmetric surfaces in \cite{ka08}.

\begin{figure}[t]

\includegraphics[width=8cm,angle=0]{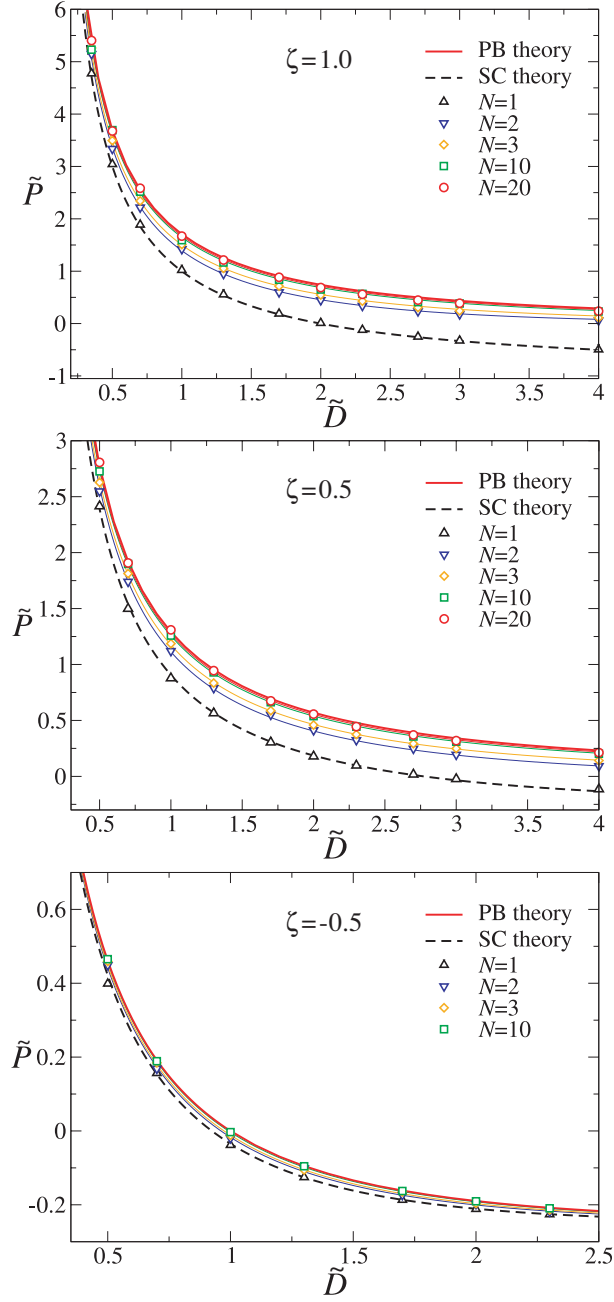}
\caption{(Color online) Rescaled interaction pressure, $\tilde p$, as a function of the rescaled distance, 
$\tilde L$, between charged boundaries in the 1D system of counterions. We show the results
 for three different values of the asymmetry parameter $\zeta=1.0, 0.5$,  and $-0.5$ (top to bottom).
Thick solid lines represent the PB prediction, Eqs. (\ref{pres-pb-rep}) and (\ref{pres-pb-att}), and the dashed lines 
are the SC prediction, Eq. (\ref{eq:P_SC}). Symbols correspond to MC simulations data and thin solid lines are the 
exact results, Eq. (\ref{eq:P_exact}), at different number of counterions as indicated in the graphs. }
\label{fig:sim_P}
\end{figure}

\section{Numerical simulations}

We next consider Monte-Carlo simulations of the 1D system of counterions treated in the preceding sections 
using both exact as well as approximate (limiting) analytical approaches. Monte-Carlo simulations enable us 
to access the parameter space inaccessible to the limiting WC and SC theories, and thus may be compared 
directly with the exact solution presented in Section \ref{sec:exact}.

We proceed by simulating a system of $N$ counterions in a finite interval $x\in[0, L]$ in the canonical 
ensemble by applying a standard Metropolis algorithm.  The Hamiltonian of the system is defined in Eq. 
(\ref{eq:Hamilt}) and the electroneutrality condition is imposed {\em via} Eq. (\ref{eq:neutral}). We run the 
simulations by up to $10^8-10^9$ Monte-Carlo steps per particle with $10^7-10^8$ steps used for relaxation 
purposes.

The interaction pressure between the two charged boundaries is calculated {\em via} the contact-value
theorem (see also Eq. (\ref{eq:contact_theorem}))
\begin{equation}
\beta p= \rho(0) - \frac{\beta \sigma_1^2}{ 2 } = \rho(L) - \frac{\beta \sigma_2^2}{ 2 },
\end{equation}
where $\rho(0)$ and $\rho(L)$ 
represent the counterion density at contact with the first ($\sigma_1$) and the 
second ($\sigma_2$)  plate, respectively. The resulting pressure computed from the contact density at either 
surface is found to be the same within the numerical errorbars. Simulations were conducted in rescaled units 
for various number of counterions $N=1,\ldots,20$ and for the asymmetry parameters $\zeta=-0.5, 0.5$ and 
1.0. The coupling parameter in each case follows from Eq. (\ref{eq:rescl_neutral}).

The simulated pressure is shown in rescaled units in Fig. \ref{fig:sim_P} (symbols) for the rescaled 
pressure $\tilde p={\beta p\ell_{\mathrm{B}}}/{(\sigma_1/e)^2}$ as a function of the rescaled 
distance, $\tilde L=L/\mu$, between the charged boundaries. As seen, the interaction pressure decays 
monotonically with separation distance $L$ for all values of $\zeta$ and $N$. Also in all cases, the 
simulation data (symbols) and the exact results (thin solid curves) are nicely bracketed by the mean-field 
PB result (thick solid curve) and the SC result (dashed curve).

For the symmetric case $\zeta=1$, the simulation results are spot on the SC curve (dashed line) for $N=1$ 
and approach slowly the attractive asymptotic SC pressure at large separations $\tilde L\rightarrow\infty$, 
that is $\tilde p^{\mathrm{SC}}_\infty = -\zeta^2$. As the number of counterions increases the pressure 
becomes less attractive and eventually for sufficiently large $N$, the simulation data tend to the PB curve 
(thick solid curve) exhibiting repulsive pressure at all separations. This is because for large $N$ the 
system effectively splits into two nearly electroneutral halves, where each bounding surface is neutralized 
by its corresponding layer of counterions. This also indicates that the entropic contribution from 
counterions which favors repulsion becomes important at large $N$. For finite $N$, the simulation data is 
described neither by the PB theory nor by the SC theory. In this case, an excellent agreement is found 
between the data and the exact results given by Eq. (\ref{eq:P_exact}) (shown by thin solid lines).

Similar trends are observed for asymmetric systems as shown for $\zeta=0.5$ and $\zeta=-0.5$ in the figure. In the 
case of oppositely charged boundaries ($\zeta<0$), it turns out that the PB and SC curves and hence the exact results 
roughly coincide and become less distinguishable. A reasonable explanation for this would be that for oppositely 
charged surfaces the counterions mostly feel the effect of the strong uniform external field provided by the surface 
charges, which acts similarly in the strong as well as the weak-coupling limit. Thus the mean-field and the 
strong-coupling approaches should converge. For small external field, as in the case of similarly charged surfaces, 
the mean-field theory depends more on the local counterion density whereas the strongly coupled counterions still feel 
mostly the external field. Thus the difference between the WC and the SC frameworks in the $\zeta<0$ and $\zeta>0$ 
cases. We also emphasize that the above discussion holds in rescaled representation as the pressures are plotted here 
in rescaled units; in actual units, Fig. \ref{fig:sim_P} corresponds to different ranges of separation, $L$, for the 
WC and SC regimes as the Gouy-Chapman length, $\mu$, is typically very different between the two limits, {\em i.e.}, 
it would be small at high couplings and large at small couplings as may be realized, {\em e.g.}, by changing the 
counterion valency at fixed surface charge densities and Bjerrum length.

Note that for oppositely charged boundaries, both the PB and the SC pressure can be attractive at large 
separations, and hence the pressure at any finite $N$ can be attractive. It is also notable that for charged 
surfaces of opposite sign, the PB analysis in general performs much better than for the surfaces of equal 
sign and that the asymptotic SC value is approached more quickly.

In all cases considered here the theoretical and the simulated values of the interaction pressure converge 
for very small surface-surface separations. In fact, the SC and PB results coincide in the leading order as 
the rescaled distance, $\tilde L$, tends to zero, since both are dominated by the osmotic pressure of 
counterions. One thus finds
\begin{equation}
\tilde p(\tilde L)\simeq \frac{1+\zeta}{\tilde L}\qquad \tilde L\ll 1, 
\label{rep-limit1}
\end{equation}
which is of course nothing but the ideal-gas osmotic pressure of counterion confined between the two plates 
({\em i.e.}, $p=N k_{\mathrm{B}}T/L$ in actual units).

\section{Conclusions}

We have analyzed the statistical physics of an overall electroneutral system, of one-dimensional counterions confined 
between two charged surfaces. This can be seen as a simple model for charged lipid multilayers or stiff mica sheets 
neutralized by counterions and confined between external charged plates. The thermodynamics can be solved exactly and 
a widely varying behavior of effective interaction between the confining plates can be discerned.

Apart from performing exact statistical mechanics and MC simulations, we have also analyzed these systems within the 
mean-field (Poisson-Boltzmann) or weak-coupling approximation and within the strong-coupling approximation. These 
approximations are in fact independent of the dimensionality of the problem and it is therefore interesting to see how 
they compare with our exact results in one dimension. For a large number of particles it is found that the mean-field 
approximation works well. This is physically understandable as the counterion distribution can be reasonably assumed 
to have a continuous profile as predicted by the mean-field theory and also the inter-particle interaction in one 
dimension is long range. In the case of few counterions it works less well as the effect of correlations becomes 
important.  In this regime the SC approximation will perform better as it takes into account properly the discreteness 
of the counterions charges. Indeed, as the strong-coupling expansion is basically a form of the virial expansion it is 
to be expected that it works better in the case of few counterions.

On the level of the approximate evaluations of the partition function, only the fluctuation contribution to the 
weak-coupling limit depends on the dimensionality of the problem because of fluctuations in the plane of the film in 3D 
that are not present in 1D for which the summation over the in-plane modes is absent. In particular, the fluctuation 
contribution to the interaction pressure scales with the distance $L$ as $p_2 \sim -\Xi_{\mathrm{1D}} /L$ in 1D and as 
$p_2 \sim - \Xi_{\mathrm{3D}} \ln L/L^3$ in 3D and for same-sign surfaces ($\zeta\geq 0$) at sufficiently large 
separations. The fluctuational contribution to the pressure is always attractive, reflecting the fact that 
electrostatic correlations mediated by counterions always favor attraction between the charged boundaries. Since the 
coupling parameter in 1D, $\Xi = \Xi_{\mathrm{1D}}$, depends inversely on the number of counterions, the fluctuation 
contribution to the interaction pressure that scales linearly with $\Xi$ is in this case vanishingly small, a 
situation clearly confirmed by exact and MC evaluation of the partition function. We can solve for $p_2(L)$ exactly
in 1D and 3D and evaluate the relative importance of the contributions from fluctuations and from mean field theory.
For example, in 1D with $N=20,\;\zeta=1$ ($\Xi_{\mathrm{1D}}=0.1$), we find $p_2/p_0 \lnsim 0.1$ for $\tilde{L} < 5$;
whereas in 3D we have $p_2/p_0 \lnsim 0.1$ for $\tilde{L} > 0.05$ when $\Xi_{\mathrm{3D}} = 0.1$, and $\tilde{L} > 20$ 
when $\Xi_{\mathrm{3D}} = 1.0$. This confirms the results inferred from the asymptotic behaviour of $p_2(L)$.

In the case of symmetric surface charges the mean-field predictions always give a positive (repulsive)  pressure 
between the two plates.  However the SC and exact results do predict a possible attraction at large inter-plate 
separations. This effect is basically due to strong correlations arising when the counterions cannot neutralize both 
surface charges simultaneously and thus the residual surface charges are both attracted to the residual counterion 
charge in between the two plates. As the number of counterions increases this effect disappears and the system 
exhibits a more mean-field-like behavior.

As regards future work it would be interesting to see if the one-dimensional theory can be adapted to model a full 
three dimensional system. There is some hope that a proper reformulation of the statistical mechanics of this system, 
by isolating explicitly the dependence of the local fields in the transverse as opposed to the longitudinal direction 
with respect to the bounding surface normals, might open up new prospects for approximations that would work also in 
the gray area where the WC and the SC fail. This approach may work due to the fact that the mean-field and 
strong-coupling approximations are dimensionality independent. Whether or not it can be expected to work will depend 
on whether charge correlations in the direction perpendicular to the film dominate the physics or whether lateral 
correlations (such as the formation of locally two dimensional Wigner-crystal-like structures in the plane of the 
film) dominate.

\begin{acknowledgments} This research was supported in part by the National Science Foundation under Grant No. 
PHY05-51164 (while at the KITP program {\em The theory and practice of fluctuation induced interactions}, UCSB, 2008). 
D.S.D acknowledges support from the Institut Universtaire de France. R.P. would like to acknowledge the financial 
support by the Agency for Research and Development of Slovenia, Grants No. P1-0055C, No. Z1-7171, No. L2- 7080. This 
study was supported in part by the Intramural Research Program of the NIH, National Institute of Child Health and 
Human Development.

\end{acknowledgments}

\end{document}